\documentclass{jpp}
\usepackage{graphicx}
\usepackage[hidelinks]{hyperref}

\newcommand{\Nfp}{N_{\mathrm{p}}}
\newcommand{\Wr}{\mathrm{Wr}}
\newcommand{\Tw}{\mathrm{Tw}}
\newcommand{\Lk}{\mathrm{Lk}}

\usepackage{bm}

\usepackage[utf8]{inputenc}
\usepackage[T1]{fontenc}
\usepackage{amsmath}

\shorttitle{Controlling writhe in stellarator optimization}

\shortauthor{T. M. Schuett, and S. A. Henneberg}

\title{Controlling writhe in stellarator optimization}

\author{Tobias M. Schuett \corresp{\email{tobims@mit.edu}} \and Sophia A. Henneberg}

\affiliation{Plasma Science and Fusion Center, Massachusetts Institute of Technology, Cambridge, MA, USA}

\begin{document}

\maketitle

\begin{abstract}
 The writhe of a closed curve measures its coiling, and for a ribbon along it, writhe is the contribution to the linking number not due to twist.
In stellarators, large writhe implies large helical excursion and has previously been either an uncontrolled outcome in optimization, or fixed beforehand by the choice of the coordinate system. We present a simple method to target the underlying writhe of the plasma boundary during optimization. 
Our method calculates writhe and linking number directly from the underlying elliptical boundary of the stellarator. 
It therefore does not require any properties of the MHD equilibrium solve, such as the true magnetic axis, making it fast to evaluate and smooth to differentiate. 
Examples for quasi-helically symmetric stellarators show that this target works well in practice.

\end{abstract}

\section{Introduction}

In contrast to tokamaks, stellarators are inherently three-dimensional and can therefore feature a non-zero \textit{linking number} associated with the plasma boundary or its magnetic axis.
The C\u{a}lug\u{a}reanu--White--Fuller theorem \citep{calugareanu1959integrale, cualuguareanu1961classes, White1969, fuller1971writhing, Moffatt_Ricca_1992}
\begin{equation}\label{eq:CWF_theorem}
    \Lk = \Wr + \Tw
\end{equation}
states that the linking number $\Lk$ of a ribbon associated with a closed space curve can be decomposed into writhe $\Wr$ and twist $\Tw$. Put in simple terms, writhe measures the coiling of a curve in space.
While $\Wr$ is a property of the closed space curve alone, both the twist $\Tw$ and thus the linking number $\Lk$ depend on the frame that defines the ribbon, and different choices can be made.
The \textit{self-linking} refers specifically to the linking number in the Frenet frame, and therefore counts the number of rotations the curve's normal vector completes over one toroidal transit. In this frame the twist then equals the net torsion of the curve.

For a stellarator, it is natural to define its associated closed space curve as the magnetic axis, or an approximation thereof.
When considering the possible contributions to the rotational transform close to the magnetic axis in the Frenet frame, one finds that these are net toroidal current (as in a tokamak), net rotation of the poloidal cross-sections, net torsion of the magnetic axis, or self-linking of the magnetic axis \citep{Mercier_1964, Helander_2014}.
It was pointed out in \citet{Pfefferle_2018} that even though writhe does not feature explicitly in Mercier's formula, it does so indirectly through the difference of self-linking and integrated torsion, and it was argued that writhe is the physically relevant quantity for the on-axis rotational transform due to the non-planar geometry of the axis.

Most of today's optimized stellarators feature net rotation of cross-sections and integrated torsion, but little writhe, as is evident from their relative planarity.
The first stellarator ever conceived, however, was shaped like a figure eight and therefore featured large writhe \citep{spitzer1958stellarator}.
Recently, there has been renewed interest in stellarators with increased writhe due to their ability to be supported by more planar coils \citep{Plunk_2025, Salm2026, rodriguez2026_arxiv}. This has motivated the development of a flexible coordinate frame in \texttt{GVEC} \citep{hindenlang2026gvec}.

While stellarators are able to generate rotational transform through 3D shaping, this shaping must be carefully tailored to ensure the collisionless confinement of trapped particles. One such possibility is to make the field omnigenous. The two most common strategies to make the field omnigenous are quasi-symmetry (QS) and quasi-isodynamicity (QI) \citep{Helander_2014}. Analytic constructions of equilibria via an asymptotic expansion near the magnetic axis have shown that omnigenous equilibria with large writhe exist, both for QS \citep{landreman2022mapping} and QI \citep{Rodriguez_Plunk_2026}. It should be noted that QS equilibria with large writhe appear limited to those with quasi-helical symmetry (QH), i.e., those where the magnetic field strength on each flux surface depends only on a single helical angle in Boozer coordinates \citep{landreman2022mapping}.

While near-axis construction is a useful tool to map and understand the parameter space of available configurations, numerical optimization is required once various design criteria resulting from different physics must be balanced against each other to obtain a more mature stellarator design. 
In stellarator optimization, writhe has previously either been an uncontrolled outcome of the optimization or has been set beforehand by the choice of the initial condition and its coordinate frame, such as in the \texttt{GVEC} code \citep{hindenlang2026gvec}.
Moreover, optimizations performed in cylindrical coordinates tend to produce configurations with small writhe, such as the precise QH configurations of \citet{LandremanPaul2022}.
Given its apparent correlation with coil planarity \citep{Plunk_2025,Salm2026,rodriguez2026_arxiv}, it is undesirable for the writhe to be an uncontrolled outcome of the optimization or to be fixed prior to the optimization. Instead, one would like to trade writhe off against other targets during the optimization.

The goal of this paper is to present a simple method to control the writhe during stellarator optimization. Rather than using the self-linking of the axis, our method calculates the writhe of a configuration from the shape of the trial plasma boundary during the optimization, and can thus be evaluated without the solution of the MHD equilibrium. We make this choice as we seek to control the contribution of writhe to the rotation of the plasma boundary, and not its contribution to the rotation of the axis normal. Conveniently, this choice also circumvents issues with the Frenet frame, which is ill-defined at points of zero curvature. 
Instead of the true magnetic axis, which would require the MHD equilibrium solution, we use the centerline of the plasma boundary to calculate the writhe for each trial boundary during the optimization. By using the underlying ellipse of the plasma boundary and projecting it onto the plane perpendicular to its centerline, the linking number associated with the plasma boundary can also be calculated in a straightforward manner.

The paper is structured as follows. In section \ref{sec:writhe} expressions for the writhe and linking number based on the underlying ellipse and centerline of the plasma boundary are derived. Section \ref{sec:optimization} describes the optimization method to target configurations with a specific writhe. In section \ref{sec:results} we show results of writhe-optimized QH configurations with two, three, and four field periods. Finally, we conclude in section \ref{sec:discussion}.

\section{Writhe and linking number}\label{sec:writhe}
\subsection{Writhe}
We seek to calculate the writhe of the centerline of the stellarator boundary, as a proxy for its true magnetic axis. As commonly done during stellarator optimization, we represent the boundary of the stellarator as a double Fourier series
\begin{equation} \label{eq:rz-fourier-boundary}
    \begin{aligned}
        R(\vartheta, \varphi) &= \sum_{m=1}^{m_\mathrm{max}} \sum_{n=-n_\mathrm{max}}^{n_\mathrm{max}} R_{m,n} \cos(m\vartheta - \Nfp n \varphi) + \sum_{n=0}^{n_\mathrm{max}} R_{0,n} \cos(\Nfp n \varphi), \\
        Z(\vartheta, \varphi) &= \sum_{m=1}^{m_\mathrm{max}} \sum_{n=-n_\mathrm{max}}^{n_\mathrm{max}} Z_{m,n} \sin(m\vartheta - \Nfp n \varphi) - \sum_{n=1}^{n_\mathrm{max}} Z_{0,n} \sin(\Nfp n \varphi),
    \end{aligned}
\end{equation}
where we have assumed stellarator symmetry \citep{stellarator_symmetry}. For the following, we define the centerline of the stellarator as the closed curve obtained by retaining only modes with $m=0$,
\begin{equation}
    R_0(\varphi) = \sum_{n=0}^{n_\mathrm{max}} R_{0,n} \cos(\Nfp n \varphi) \quad\text{and}\quad
    Z_0(\varphi) = - \sum_{n=1}^{n_\mathrm{max}} Z_{0,n} \sin(\Nfp n \varphi).
\end{equation}
This gives the closed space curve
\begin{equation}
    \mathbf{x}(\varphi) = R_0(\varphi) \mathbf{e}_R + Z_0(\varphi)\mathbf{e}_Z,
\end{equation}
where $\mathbf{e}_R$ and $\mathbf{e}_Z$ are the standard cylindrical unit vectors.
It serves as a proxy for the true magnetic axis, and is by construction the centerline of the underlying elliptical boundary shape of the stellarator that we will use in the next section to calculate the linking number.

Using $\mathrm{d}\mathbf{e}_R/\mathrm{d}\varphi = \mathbf{e}_\varphi$ and $\mathrm{d}\mathbf{e}_\varphi/\mathrm{d}\varphi = -\mathbf{e}_R$, the centerline's tangent and its derivative are
\begin{equation}
  \bm t = \frac{\bm x'}{|\bm x'|} \qquad\text{and}\qquad
  \bm t' = \frac{\bm x'' - (\bm x''\cdot\bm t)\,\bm t}{|\bm x'|},
\end{equation}
where primes denote $\mathrm{d}/\mathrm{d}\varphi$.
Generally, the writhe of the closed curve $\bm x$ is the Gauss double integral \citep{calugareanu1959integrale}
\begin{equation}
  \Wr = \frac{1}{4\pi} \oint\!\!\oint
    \frac{\big(\bm x'(\varphi)\times\bm x'(\varphi')\big)\cdot\big(\bm x(\varphi)-\bm x(\varphi')\big)}
         {|\bm x(\varphi)-\bm x(\varphi')|^{3}}\,\mathrm{d}\varphi\,\mathrm{d}\varphi' .
\end{equation}
Instead of evaluating it directly, we use Fuller's theorem \citep{fuller1971writhing, fuller1978decomposition}, which says that the difference in writhe between two closed curves $C_0$ and $C_1$ is given by
\begin{equation}
  \Wr(C_1) - \Wr(C_0) = \frac{1}{2\pi}\oint
    \frac{(\bm t_0\times\bm t_1)\cdot(\bm t_0' + \bm t_1')}{1 + \bm t_0\cdot\bm t_1}\,\mathrm{d}\varphi .
  \label{eq:fuller}
\end{equation}
Here $\bm t_0$ and $\bm t_1$ are the tangent vectors of $C_0$ and $C_1$, respectively. Equation \eqref{eq:fuller} holds as long as $C_1$ can be continuously deformed into $C_0$ through closed curves that neither self-intersect nor feature a tangent vector that is antiparallel to $\bm t_0$ at any point. We let $C_1$ represent the centerline of our stellarator and define $C_0$ to be a planar circle $R=\text{const}$, $Z=0$,
parametrized by the same angle $\varphi$.
The centerline of typical stellarators crosses each poloidal plane only once, so it cannot self-intersect, and since $\bm t\cdot\mathbf{e}_\varphi > 0$ the tangent of the intermediate curves will never be antiparallel to $\bm t_0$.
Noting that the writhe of our reference curve $C_0$ vanishes, we get
\begin{equation}
  \Wr(C_1) = \frac{1}{2\pi}\oint
    \frac{(\mathbf{e}_\varphi\times\bm t)\cdot(\bm t' - \mathbf{e}_R)}{1 + \mathbf{e}_\varphi\cdot\bm t}\,\mathrm{d}\varphi,
  \label{eq:writhe}
\end{equation}
where we used $\bm t_0 = \mathbf{e}_\varphi$.

\subsection{Linking number}
In this section we calculate the linking number associated with the plasma boundary by defining a ribbon based on the major axis of the underlying elliptical boundary. The writhe is linked to the twist and linking number of a ribbon through the C\u{a}lug\u{a}reanu--White--Fuller theorem, equation \eqref{eq:CWF_theorem}. When considering writhe values to target during the optimization, it appears natural to do so relative to the linking number of the ribbon associated with the plasma boundary, since a large fraction of writhe then corresponds to a minimization of twist for this ribbon. We thus seek a simple method of evaluating the linking number associated with the plasma boundary.

Keeping only the modes with $m \le 1$, the boundary cross-section in each
plane $\varphi = \text{const}$ becomes an ellipse centered on the centerline,
\begin{equation}
  \begin{pmatrix} R \\ Z \end{pmatrix}
  = \begin{pmatrix} R_0(\varphi) \\ Z_0(\varphi) \end{pmatrix}
  + \mathsf S(\varphi) \begin{pmatrix} \cos\vartheta \\ \sin\vartheta \end{pmatrix},
  \qquad
  \mathsf S = \begin{pmatrix} S_{RC} & S_{RS} \\ S_{ZC} & S_{ZS} \end{pmatrix},
\end{equation}
with
\begin{align}
  S_{RC} &= \sum_{n} R_{1n}\cos(n\Nfp\varphi), &
  S_{RS} &= \sum_{n} R_{1n}\sin(n\Nfp\varphi), \\
  S_{ZC} &= -\sum_{n} Z_{1n}\sin(n\Nfp\varphi), &
  S_{ZS} &= \sum_{n} Z_{1n}\cos(n\Nfp\varphi).
\end{align}

Depending on how non-planar the stellarator boundary is, the centerline will cross the poloidal planes at an oblique angle. For a strongly inclined centerline the cut can therefore be elongated even where the actual cross-section of the plasma is nearly circular. Its orientation then follows the inclination of the centerline rather than the shape of the plasma. We therefore project the cross-section at constant $\varphi$ along $\bm t$ onto the plane perpendicular to the centerline, and use this projected ellipse to calculate both the elongation and the linking number.

In this plane we use the orthonormal basis vectors
\begin{equation}
  \bm f_1 = \frac{\mathbf{e}_R - (\mathbf{e}_R\cdot\bm t)\,\bm t}{|\mathbf{e}_R - (\mathbf{e}_R\cdot\bm t)\,\bm t|}
  \qquad\text{and}\qquad
  \bm f_2 = \bm t\times\bm f_1 .
  \label{eq:frame}
\end{equation}
These are well defined because $\bm t\cdot\mathbf{e}_\varphi > 0$, so $\mathbf{e}_R$ is never parallel to $\bm t$.\footnote{It should be noted that very exotic shapes, such as knotted configurations, are excluded by making this assumption.}
We write the columns of $\mathsf S$ as 
\begin{equation}
    \begin{aligned}
    \bm c_C &= S_{RC}\,\mathbf{e}_R + S_{ZC}\,\mathbf{e}_Z, \\
    \bm c_S &= S_{RS}\,\mathbf{e}_R + S_{ZS}\,\mathbf{e}_Z,
    \end{aligned}
\end{equation}
and build the shape matrix of the ellipse in the new frame by projecting $\bm c_C$ and $\bm c_S$ onto the basis vectors of the new frame,
\begin{equation}
    \mathsf S_\perp
  = \begin{pmatrix} \bm f_1\cdot\bm c_C & \bm f_1\cdot\bm c_S \\
                    \bm f_2\cdot\bm c_C & \bm f_2\cdot\bm c_S \end{pmatrix}
  = \begin{pmatrix} S_{1C} & S_{1S} \\ S_{2C} & S_{2S} \end{pmatrix}.
\end{equation}
Then the cross-sections within the new perpendicular frame are given by
\begin{equation}
    \bm x(\varphi) + \big(\bm f_1\;\; \bm f_2\big)\,\mathsf S_\perp(\varphi)
    \begin{pmatrix} \cos\vartheta \\ \sin\vartheta \end{pmatrix}.
\end{equation}
The squared semi-axes of the ellipse are given by the eigenvalues $\lambda_+ (\varphi) \ge \lambda_- (\varphi)$ of the symmetric matrix $\mathsf Q = \mathsf S_\perp\mathsf S_\perp^{\mathsf T}$. We define the elongation as 
\begin{equation}
    E(\varphi) = \sqrt{\lambda_+/\lambda_-}.
\end{equation}

To obtain the linking number we note that the angle $\alpha$ between the major axis and $\bm f_1$ is given by
\begin{equation}
  2\alpha = \arg\!\big[(Q_{11} - Q_{22}) + 2\mathrm{i}\,Q_{12}\big],
  \label{eq:alpha}
\end{equation}
where $Q_{11}$, $Q_{22}$ and $Q_{12}$ are the components of $\mathsf Q$ \citep{korn_korn_2000_mathematical_handbook}.
Since the major axis is a line rather than an oriented vector,
$\alpha$ is defined modulo $\pi$. We follow the continuous branch of $\alpha(\varphi)$
around the torus.
The linking number follows directly from the rotation of the major axis,
\begin{equation}
  \Lk = \frac{\alpha(2\pi) - \alpha(0)}{2\pi}.
  \label{eq:lk}
\end{equation}
It is thus the accumulated rotation of the ribbon defined by the major axis of the underlying ellipse of the stellarator. Since this ribbon has to close onto itself for a stellarator boundary, it is either an integer or a half-integer. Integer values correspond to the case where this ribbon has the same orientation after one toroidal turn, half-integer to the cases where the ribbon closes onto itself with its edges flipped.

\section{Optimization}\label{sec:optimization}
\subsection{Objective}
We are interested in finding quasi-helically symmetric stellarators with varying writhe via numerical optimization. To this end we minimize the objective
\begin{equation}\label{eq:target}
    f = w_\mathrm{A} (A-A^*)^2 + f_\mathrm{QH} + w_\mathrm{Wr} (\Wr - \Wr^*)^2 + f_\mathrm{E}.
\end{equation}
Here $A^*=8$ is the desired aspect ratio (following the \texttt{VMEC} definition) and $f_\mathrm{QH}$ is the two-term metric for quasi-helical symmetry \citep{LandremanPaul2022},
\begin{equation}\label{eq:QH-objective}
    f_\mathrm{QH} = \sum_{s_j} w_j \Big\langle\big(\frac{1}{B^3}[(N-\iota M)\mathbf{B}\times \nabla B \cdot \nabla\psi - (G+NI)\mathbf{B}\cdot\nabla B]\big)^2\Big \rangle.
\end{equation}
Here $B$ is the magnetic field strength, $N=-\Nfp$ and $M=1$ specify the quasi-symmetry helicity, $\iota$ is the rotational transform, $G$ is $\mu_0/(2\pi)$ times the poloidal current outside of surface $s$, $I$ is $\mu_0/(2\pi)$ times the toroidal current inside surface $s$, where $s$ is the normalized toroidal flux. The sum runs over the eleven surfaces $s_j = 0, 0.1, \dots, 1$, and we set all weights $w_j = 1$.

The third term in \eqref{eq:target} constrains the optimizer to the desired writhe, which we set relative to the stellarator's linking number,
\begin{equation}
    \Wr^* = c_\mathrm{Wr}\, \Lk_0.
\end{equation}
Here $c_\mathrm{Wr}$ is the writhe fraction, where $c_\mathrm{Wr}=1$ corresponds to zero twist, and $\Lk_0$ is the linking number of the initial condition.
The final term in \eqref{eq:target} constrains the elongation from above and below,
\begin{equation}
    f_\mathrm{E} = w_\mathrm{E} \oint \left\{\max{[0, E_\mathrm{min} - E(\varphi)]^2} + \max{[0, E(\varphi)-E_\mathrm{max}]^2}\right\} \mathrm{d}\varphi.
\end{equation}
We set $E_\mathrm{min} = 1.3$ and $E_\mathrm{max} = 8$. The upper bound prevents our solution from developing large surface-to-volume ratios. The lower bound is set to prevent the cross-sections from becoming circular anywhere. A circular cross-section anywhere is undesirable, as it would make the ellipse axes, and therefore $\Lk$, ill-defined. It would also enable a transition to a different linking number. In our optimizations we would like the linking number to remain at its initial value so as to control the family of solutions we obtain.

In all but one case, $f_\mathrm{E}$ was sufficiently strong to keep $\Lk$ at its initial value. In the optimization with the largest writhe fraction, however, the optimizer crossed the barrier formed by $f_\mathrm{E}$. For this case we therefore additionally treat any evaluation with $\Lk \neq \Lk_0$ as a failed evaluation, which assigns it a large objective value, so the trust-region algorithm rejects the step and reduces its radius.

We set the weights $w_\mathrm{A}$, $w_\mathrm{Wr}$ and $w_\mathrm{E}$ all equal to one.

\subsection{Optimization method}
The optimizations are performed with \texttt{SIMSOPT} \citep{landreman2021simsopt} and the MHD equilibria to evaluate $f_\mathrm{QH}$ are calculated with \texttt{VMEC} \citep{HirshmanWhitson1983}. For these we make the vacuum assumption.
The optimizer is `trust region reflective', the default method of \texttt{SciPy}'s nonlinear least squares \citep{virtanen2020scipy}. Gradients are calculated using forward differences. For the gradient step sizes we only set one set of values, \texttt{abs\_step = 1e-8} and \texttt{rel\_step = 1e-5}. We find our optimizations to be sufficiently robust and the resulting configurations sufficiently QH that we did not scan these parameters. Our results are therefore a lower bound on what can be achieved, since some further improvement could likely be found by scanning these parameters, or using an improved optimization scheme altogether.

All optimizations are initialized from the \emph{same} simple planar rotating ellipse at aspect ratio 8 and constant elongation of $E(\varphi)=2$. The linking number associated with this boundary is $\Lk = \Nfp/2$.
The optimization is performed in a series of iterations where boundary modes with $m,|n| \leq k$ are optimized in iteration $k \in[1,4]$.\footnote{$R_{0,0}=1$ is kept fixed as it only scales the device \citep{LandremanPaul2022}.} We make one exception to this general rule: all modes with $m\leq1$ and $|n|\leq 4$, i.e., those describing the movement of the underlying ellipse, are free to vary during \emph{all} iterations. This allows the optimizer to move the boundary to the desired writhe early on.

\section{Results}\label{sec:results}

\subsection{QH configurations of varying writhe with four field periods}
\begin{figure}
  \centering
  \includegraphics[width=\textwidth]{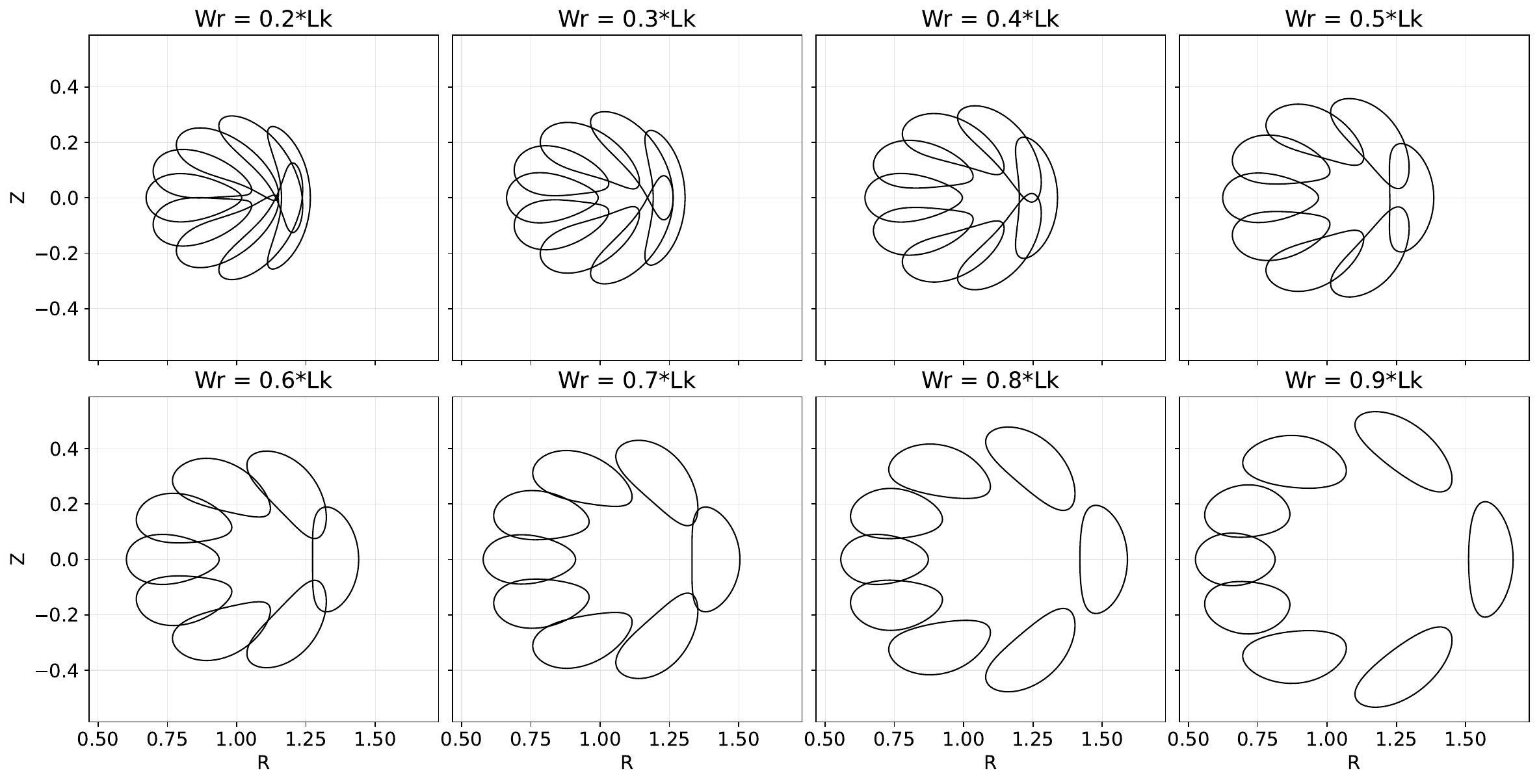}
  \caption{The cross-sections of the $\Nfp=4$ configurations for varying fractions of writhe.}
\label{fig:xsections_scan_NFP4}
\end{figure}

\begin{figure}
  \centering
  \includegraphics[width=0.9\textwidth]{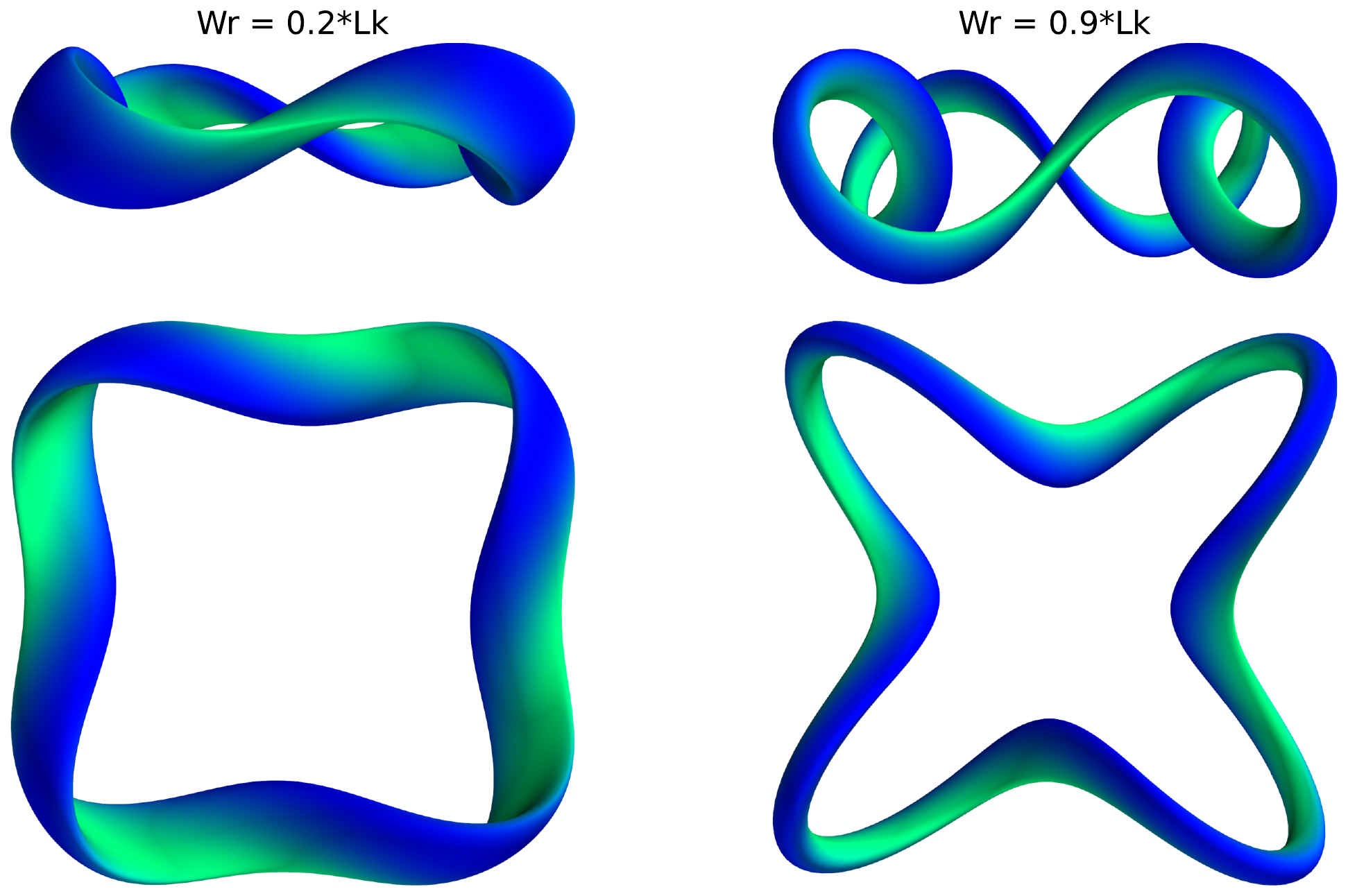}
  \caption{The 3D renditions of the $\Nfp=4$ configurations for the smallest (left) and largest (right) writhe fraction.}
\label{fig:combine_extremes_3d}
\end{figure}

\begin{figure}
  \centering
  \includegraphics[width=\textwidth]{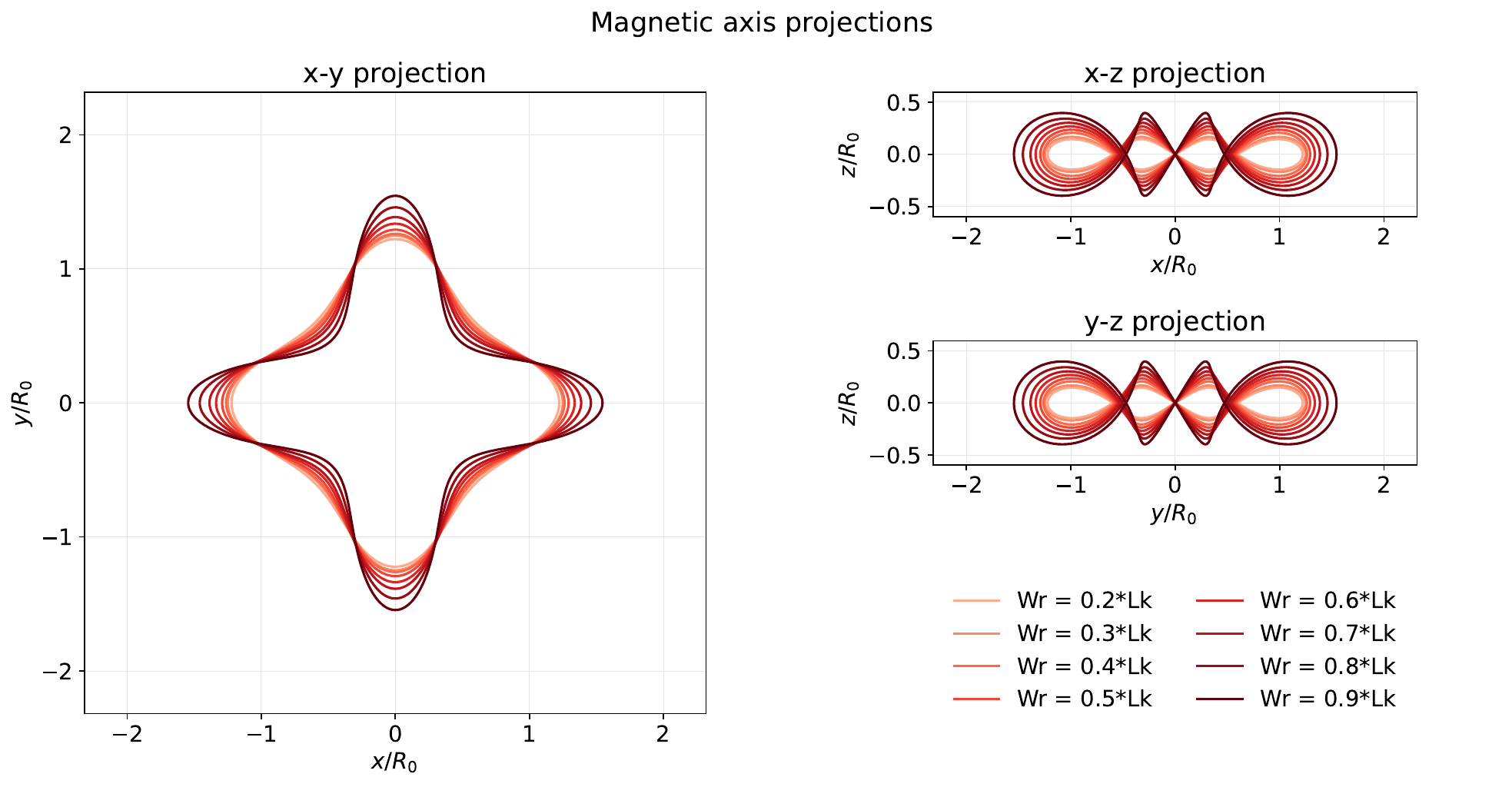}
  \caption{The projections of the centerline of the $\Nfp=4$ configurations for varying fractions of writhe.}
\label{fig:axis_projections_all}
\end{figure}

Most quasi-helical stellarator designs feature four field periods, including HSX \citep{Anderson1995} and the precise QH design from \citet{LandremanPaul2022}. We thus set $\Nfp=4$ and scan the writhe fraction $c_\mathrm{Wr}$ from 0.2 to 0.9, corresponding to $20\%$ and $90\%$ of the linking number to be produced by writhe, respectively. The final cross-sections are shown in Fig.~\ref{fig:xsections_scan_NFP4}. As the writhe fraction is increased, the configurations are more non-planar.
This is also clearly visible in Fig.~\ref{fig:combine_extremes_3d} and Fig.~\ref{fig:axis_projections_all}. Figure~\ref{fig:combine_extremes_3d} shows the plasma boundary of the two most extreme cases as a 3D rendition, while Fig.~\ref{fig:axis_projections_all} shows the projection of the centerline onto all three Cartesian planes.

\begin{figure}
  \centering
  \includegraphics[width=\textwidth]{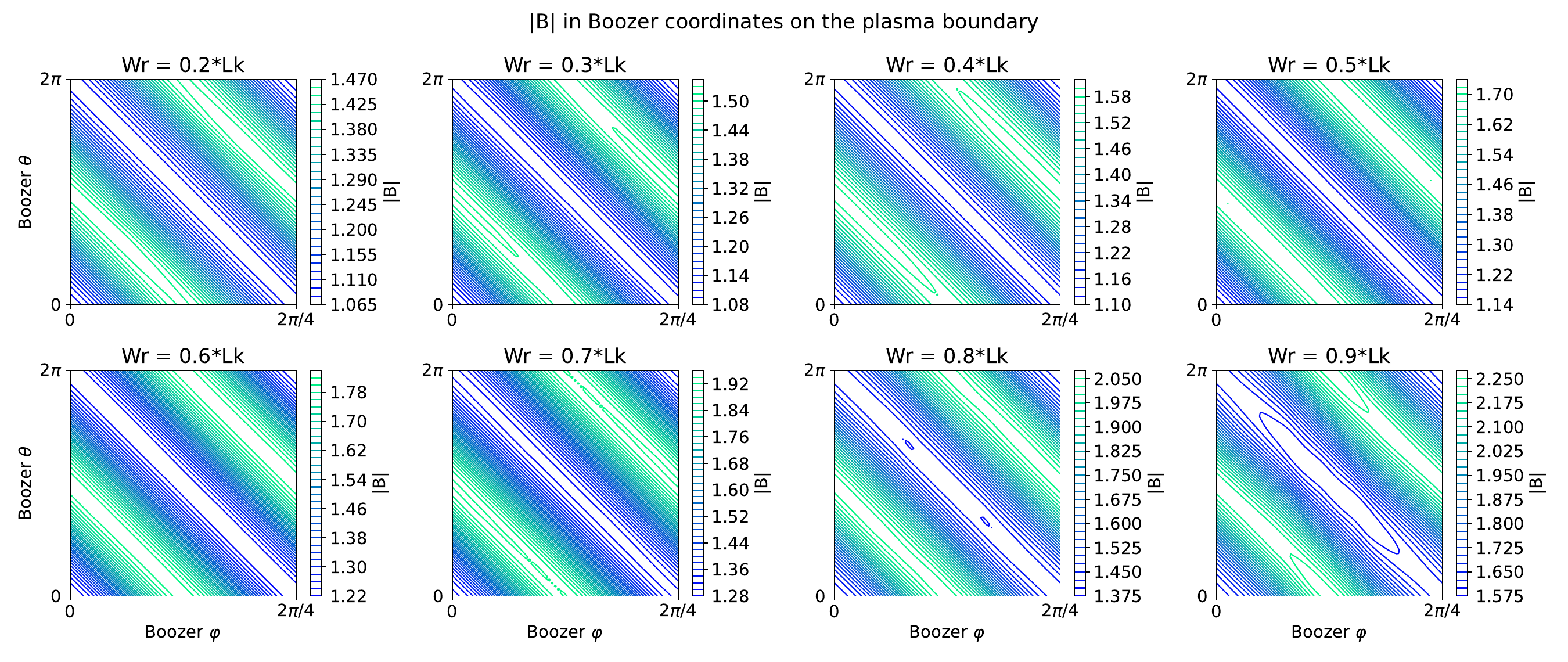}
  \caption{The contours of magnetic field strength in Boozer coordinates of the $\Nfp=4$ configurations for varying fractions of writhe.}
\label{fig:booz_all}
\end{figure}

All configurations achieve excellent QH quality as shown in Fig.~\ref{fig:booz_all}, which shows the contours of magnetic field strength in Boozer coordinates on the plasma boundary.

\begin{figure}
  \centering
  \includegraphics[width=0.6\textwidth]{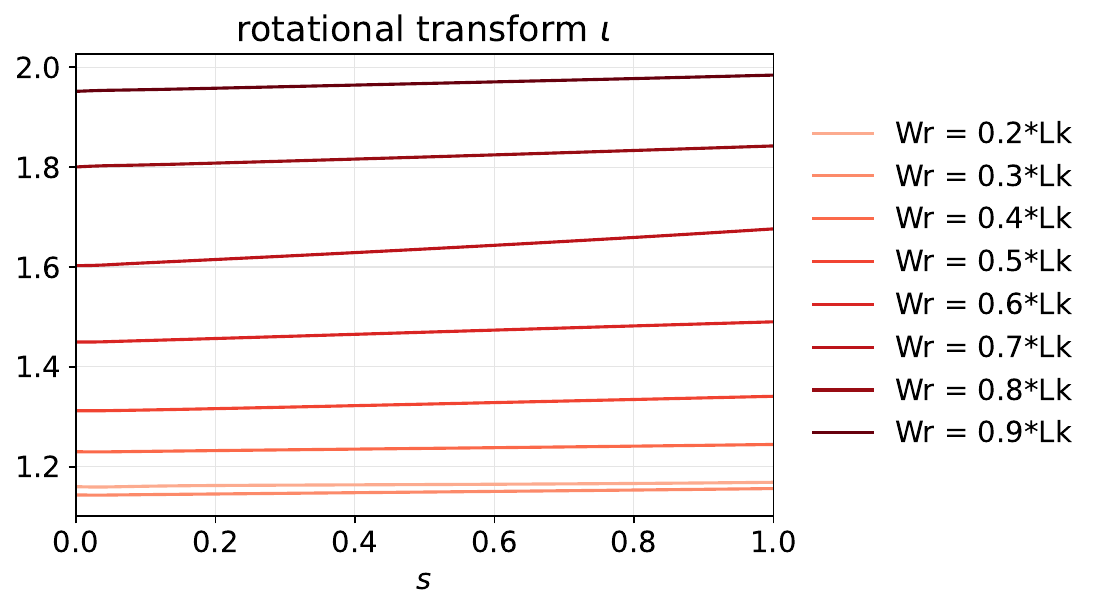}
  \caption{The rotational transform of the $\Nfp=4$ configurations for varying fractions of writhe.}
\label{fig:iota_all}
\end{figure}

The radial profiles of rotational transform are shown in Fig.~\ref{fig:iota_all}, showing a significant increase toward larger writhe, in line with
\citet{Pfefferle_2018}.

\subsection{QH with two and three field periods}
\begin{figure}
  \centering
  \includegraphics[width=0.9\textwidth]{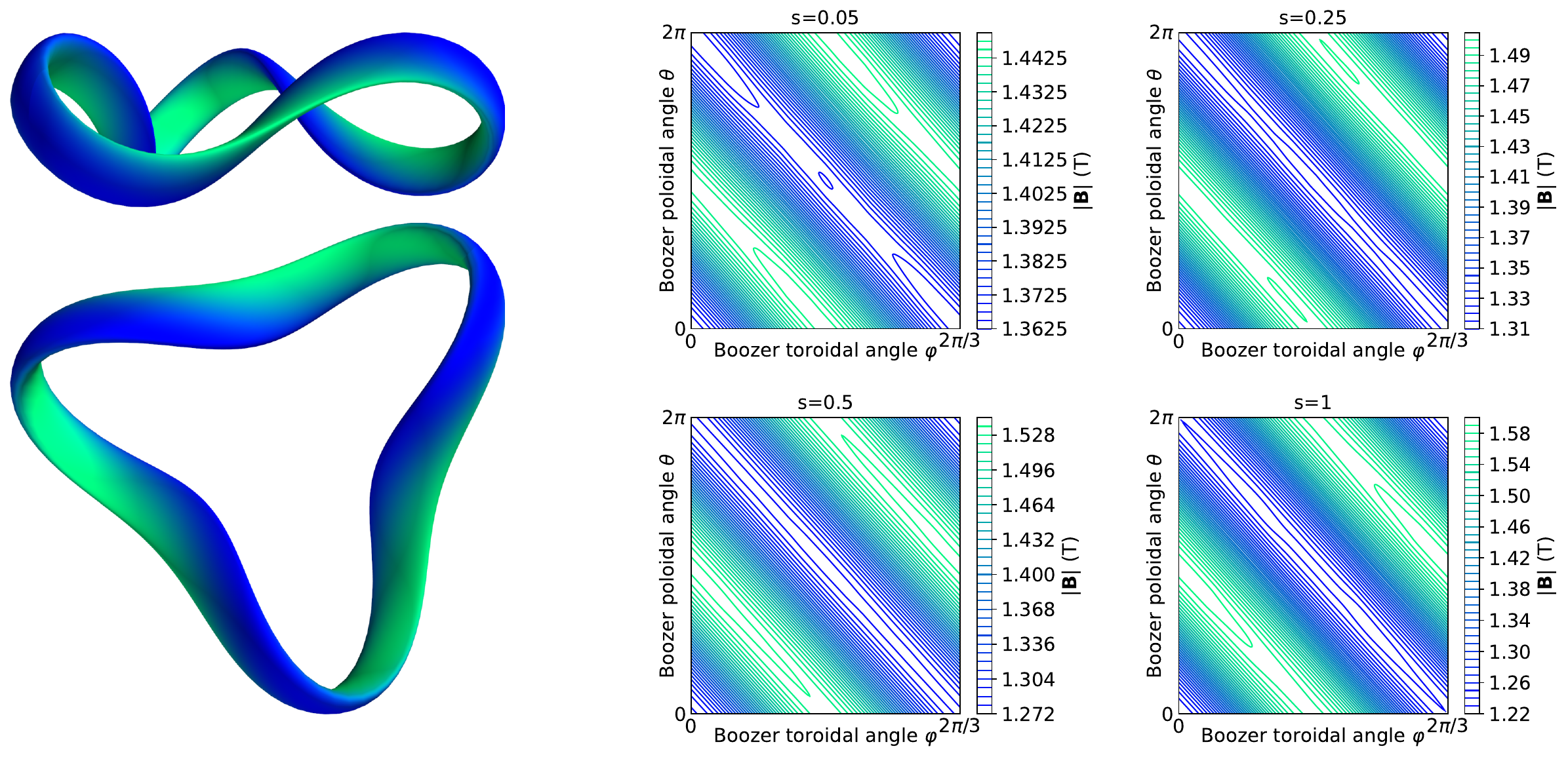}
  \caption{\textit{Three field periods --} 3D renditions (left) and contours of magnetic field strength in Boozer coordinates at four radial locations (right).}
\label{fig:combined_plot_nfp3}
\end{figure}

\begin{figure}
  \centering
  \includegraphics[width=0.9\textwidth]{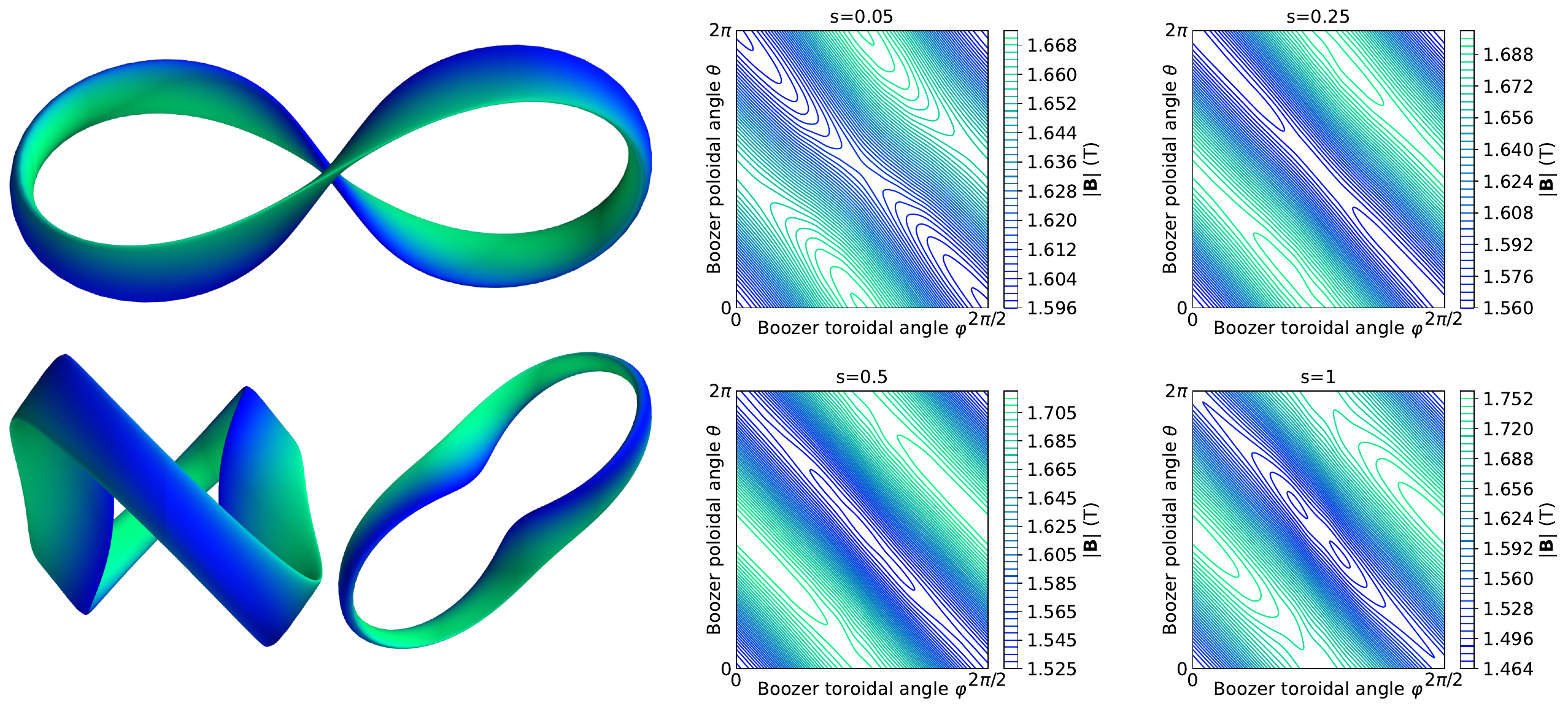}
  \caption{\textit{Two field periods --} 3D renditions (left) and contours of magnetic field strength in Boozer coordinates at four radial locations (right).}
\label{fig:combined_plot_nfp2}
\end{figure}

Our method also performs well for $\Nfp=2$ and $\Nfp=3$, which are less common choices for QH configurations. There are implicit trade-offs when choosing the number of field periods, and it is therefore useful to have a method that works robustly across different numbers of field periods. Figures~\ref{fig:combined_plot_nfp3} and~\ref{fig:combined_plot_nfp2} show the optimized equilibria with $50\%$ writhe and three and two field periods, respectively.

The configuration with two field periods resembles the figure-eight design by \citet{spitzer1958stellarator}. Such a QH configuration has previously been found from near-axis expansion \citep{landreman2022mapping} and can also be found in the QUASR database \citep{giuliani2024direct}, where such configurations arise from a large-scale global optimization of coils for quasi-symmetric vacuum fields.\footnote{See, for example, configuration 2290712 at \href{https://quasr.flatironinstitute.org/model/2290712}{\nolinkurl{quasr.flatironinstitute.org/model/2290712}}.}
However, to the knowledge of the authors a QH configuration of this type has not previously been obtained from fixed-boundary optimization.

\subsection{Confirming coil planarity}
In previous work it was shown that large writhe is associated with more planar coils \citep{Plunk_2025, Salm2026, rodriguez2026_arxiv}. 
\citet{Plunk_2025} showed that a two-field-period QI configuration that was derived from near-axis expansion and features large writhe can be reasonably realized with planar coils. \citet{Salm2026} showed that the coil planarity does indeed increase with the writhe of this family of analytic near-axis QI solutions.
\citet{rodriguez2026_arxiv} proposed solving the current potential on flux surfaces to predict coil features such as non-planarity from it.

Here we use \texttt{REGCOIL} \citep{Landreman_2017} to confirm that our configurations optimized for larger writhe do indeed feature more planar coils.
We choose the winding surface to be uniformly offset from the plasma boundary by a distance that is 1.5 times the effective minor radius.
To keep the comparison as fair as possible, we use an effective minor radius for setting this winding surface offset. We define the effective minor radius $a_\mathrm{eff}$ via the volume
\begin{equation}
    V = (\pi a_\mathrm{eff}^2) (2\pi R_\mathrm{eff}),
\end{equation}
where $R_\mathrm{eff} = L_\mathrm{axis} / (2\pi)$ is the effective major radius. Here $L_\mathrm{axis}$ is the length of the centerline of the plasma boundary, which we have used throughout this paper. The same definitions are used in the \texttt{GVEC} code \citep{hindenlang2026gvec}. 

\begin{figure}
  \centering
  \includegraphics[width=\textwidth]{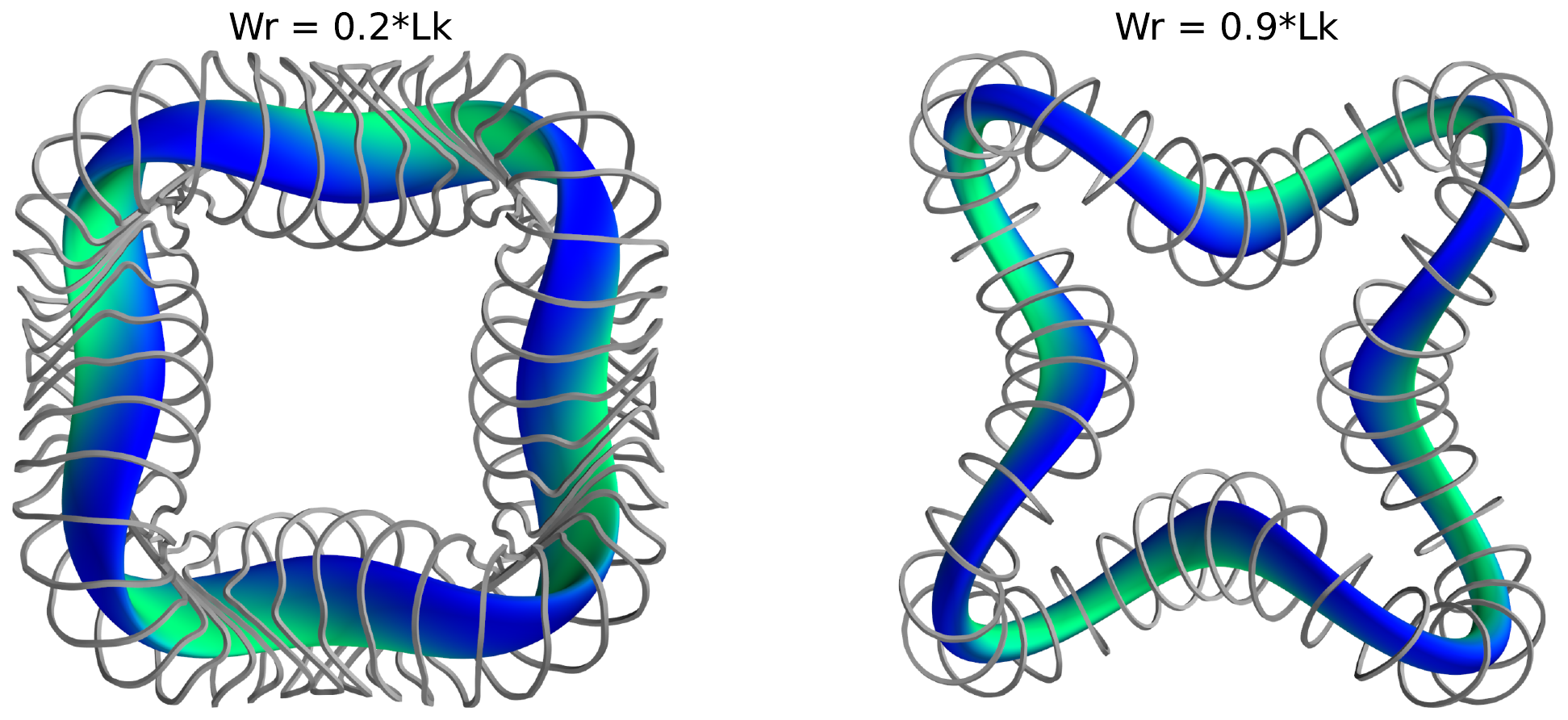}
  \caption{\textit{Confirming effectiveness of optimization --} Coils cut from \texttt{REGCOIL} for the $\Nfp=4$ configurations with the smallest (left) and largest (right) writhe considered in our optimizations.}
\label{fig:combined_coils_top}
\end{figure}

We use \texttt{REGCOIL}'s algorithm to find the regularization parameter for which the maximum relative field error $\max( |\bm B\cdot\bm n| / |\bm B| )$ equals $1\%$. The resulting current potential solution is cut into six coils per half period and is shown in Fig.~\ref{fig:combined_coils_top}. The coils are significantly more planar for the configuration with the largest writhe than for the one with the smallest writhe, confirming that our optimization had the desired effect.

\section{Conclusions}\label{sec:discussion}
In this paper we showed that a target for the writhe of the plasma boundary works well in practice during stellarator optimization. The target is calculated directly from the plasma boundary. It therefore does not rely on the MHD equilibrium solve, which makes it fast and robust to evaluate. 
Optimizations for four-field-period configurations with a range of writhe showed that the target is effective in controlling writhe in optimization. Two additional configurations with three and two field periods that feature a 50$\%$ writhe contribution to the linking number of the plasma boundary were also presented. 
All optimizations started from the same simple planar rotating ellipse, highlighting that no carefully designed initial condition is required. 
All configurations show excellent levels of quasi-symmetry. Further, it is noteworthy that these boundaries are expressed in cylindrical coordinates with Fourier modes only up to $m,|n| \leq 4$.

Here we applied the target to the optimization of quasi-helically symmetric stellarators. 
Future work could test this target for the optimization of other classes of omnigeneity. Since our method does not use the Frenet frame, it is expected to work robustly for optimizations of configurations whose magnetic axis features points of zero curvature, such as quasi-isodynamic configurations.
The calculation of the linking number was done in cylindrical coordinates in this paper but could be generalized to other coordinate systems as well, to similarly control the type of configuration while optimizing its writhe. The method presented here could therefore also be applied to optimizations using other coordinate systems. Further, the natural next step is to utilize this target in optimization to design more mature equilibria with finite plasma pressure, as well as to balance writhe against other targets such as omnigeneity, MHD stability, or constraints on the rotational transform.
 
\section*{Acknowledgements}\label{Acknowledgements}
This work was developed together with Claude Opus 5.5, which was also used to implement some of the optimization targets and optimization scripts, as well as for writing assistance in preparing the manuscript. The authors verified the work and take responsibility for its correctness.

\section*{Funding}\label{Funding}
This work was supported by the U.S. Department of Energy contract number DE-FG02-91ER-54109.

\section*{Declaration of interests}\label{Declaration_of_interest}
The authors declare no competing interests.

\bibliographystyle{jpp}
\bibliography{literature}
\end{document}